**Integrated Sensitivity Analysis of a Macroscale Hydrologic Model in the North of the Iberian Peninsula**

Patricio Yeste[1], Matilde García-Valdecasas Ojeda[1], Sonia R. Gámiz-Fortis[1], Yolanda Castro-Díez[1] and María Jesús Esteban-Parra[1]

(1) Dept. Applied Physics, University of Granada, Spain

**Corresponding author:** María Jesús Esteban-Parra (esteban@ugr.es)

## Abstract

Process-based hydrologic models allow to identify the behavior of a basin providing a mathematical description of the hydrologic processes underlying the runoff mechanisms that govern the streamflow generation. This study focuses on a macroscale application of the Variable Infiltration Capacity (VIC) model over 31 headwater subwatersheds belonging to the Duero River Basin, located in the Iberian Peninsula, through a three-part approach: (1) the calibration and validation of the VIC model for all the subwatersheds; (2) an integrated sensitivity analysis concerning the soil parameters chosen for the calibration, and (3) an assessment of equifinality and the efficiency of the calibration algorithm. The calibration and validation processes showed good results for most of the subwatersheds in a computationally efficient way using the Shuffled-Complex-Evolution algorithm. The sensitivity measures were obtained with the Standardized Regression Coefficients method through a post-process of the outputs of a Monte Carlo simulation carried out for 10 000 parameter samples for each subwatershed. This allowed to quantify the sensitivity of the water balance components to the selected parameters for the calibration and understanding the strong dependencies between them. The final assessment of the equifinality hypothesis manifested that there are many parameter samples with performances as good as the optimum, calculated using the calibration algorithm. For almost all the analyzed subwatersheds the calibration algorithm resulted efficient, reaching the optimal fit. Both the Monte Carlo simulation and the use of a calibration algorithm will be of interest for other feasible applications of the VIC model in other river basins.

# 1. Introduction

Water resources in the Mediterranean Basin have undergone dramatic changes during the 20th century as a consequence of the rising temperatures and the significant decrease of precipitation (García-Ruiz et al., 2011). The effects of climate change in this region are already noticeable and are expected to be much more pronounced by the end of the 21th century (IPCC, 2014). This fact, together with the increasing water demand for agriculture, industry and urban supply, makes the water scarcity problem of paramount importance, being its accurate identification essential for adopting adequate water management strategies and mitigation measures that ensure the sustainability of the water resources (Chavez-Jimenez et al., 2013; Garrote et al., 2016). As a part of the Mediterranean region, the Iberian Peninsula conforms a vulnerable area that has been identified as a hotspot (Diffenbaugh and Giorgi, 2012) where the streamflows have shown a marked reduction during the last half century (Lorenzo-Lacruz et al., 2012, 2013).

Being able to identify the hydrologic behavior of a basin is necessary in order to assess the effects of climate and land changing conditions, and therefore a profound description of the main hydrologic processes governing the response of the basin is required. In this way, process-based hydrologic models are powerful tools that represent the underlying runoff mechanisms governing the streamflow generation for a given basin, and therefore constitute mathematical hypotheses on how the hydrologic system functions, characterizing the potential changes of the water resources using precipitation and temperature data as inputs variables. Calibration and validation of hydrologic models are required in order to develop reliable models (Savenije, 2009), and sensitivity analysis should be carried out for a better knowledge of complex models (Song et al., 2015). Moreover, the recognition of the equifinality concept, that is, the existence of many sets of parameters conducive to good adjustments to some target observations (Beven, 2006; 2012), is unavoidable and necessary (Beven and Freer, 2001). In the context of climate modeling, these models are usually called Land-Surface Models (LSMs) and are coupled to General Circulation Models (GCMs) and Regional Climate Models (RCMs) as the land scheme that allows to simulate the biophysical processes involved in the land-atmosphere interaction (Wood et al., 2011). Although there is a subtle difference between a hydrologic model and a LSM, this distinction has become blurred over time (Clark et al., 2015).

In this respect, the Variable Infiltration Capacity (VIC) macroscale hydrologic model (Liang et al., 1994, 1996) has played the role of both LSM and hydrologic model in many previous studies. Melsen et al. (2016b) provided sufficient evidence in a meta-analysis of 192 peer-reviewed studies where the VIC model was calibrated and validated. Since its first development many efforts have been made in order to study the sensitivity of the VIC model, which has been explored in a broad sense in the following terms:

- *Sensitivity to spatio-temporal variability*: the impacts of the implemented spatial resolution in the simulated runoff and other water fluxes have been addressed in various studies, suggesting that there is a high influence of the sub-grid variability of the precipitation on the model performance (Haddeland et al., 2002; Liang et al., 2004). However, a critical spatial resolution under which a better model performance is not necessarily achieved (Liang et al., 2004) could exist. These impacts are also noticeable in calibration and validation exercises with an increase of the model accuracy at higher resolutions (Oubeidillah et al., 2014), although a high transferability of the calibrated parameters across the different resolutions may be an indicator of a poor representation of the spatial variability (Melsen et al., 2016a). Unfortunately, the time step of the calibration and validation has not kept up with the increasing spatial resolution, and this is a crucial aspect for the correct representation of the involved hydrologic processes (Melsen et al., 2016b). The fact that it is more difficult to transfer parameters across temporal resolutions than across the spatial dimension brings the need of a better representation of the spatial variability in macroscale hydrologic models (Melsen et al., 2016a).

- *Sensitivity to boundary conditions*: understanding the boundary conditions as the meteorological forcings that drive the simulations, the VIC model sensitivity to the boundary conditions has been studied through the application of different climate change scenarios and the analysis of the impacts of changing precipitation and temperature on the hydrology and water resources of several continental river basins (Nijssen et al., 2001), the Pacific Northwest (Vano et al., 2015), or more locally in the Colorado River Basin (Christensen et al., 2004; Bennet et al., 2018) and in the upper Ganga Basin (Chawla and Mujumdar, 2015). Also, these kind of studies sometimes are carried out in conjunction with other sensitivity analysis, i.e. the combined and segregated effects of climate change and land use changes on streamflow (Chawla and Mujumdar, 2015) or the parameter sensitivity under a changing climate (Bennet et al., 2018).

- *Sensitivity to initial conditions*: the question about if the hydrologic predictions are affected by the hydrologic initial conditions (i.e. the initial moisture state at the snowpack and the soil profile) or if the boundary conditions constitute the main contributor to the model simulations was studied in detail in Cosgrove et al. (2003), Wood and Lettenmaier (2009) and Li et al. (2009). It is known that the soil moisture content for the bottom soil layer of the VIC grid cells is the variable that commonly takes the longest time to reach the equilibrium, and although there is not a general agreement in how long the model spin-up period should be, since it highly depends on each particular application, it has been found that wetter states lead to faster spin-up times (Cosgrove et al., 2003; Melsen et al., 2016a). This issue is of relevance for a proper calibration and validation of the VIC model and is usually avoided by fixing a long-enough spin-up period previous to the simulation together with a wet initialization of the soil layers.

- *Sensitivity to soil and vegetation parameters*: since its generalization as a three-layer soil model (VIC-3L) in Liang et al. (1996) after its two-layer predecessor (VIC-2L, Liang et al., 1994), the sensitivity of the model to soil parameters has been studied at basin-scale and at global-scale through a cell-based approach. The basin-scale approach of Demaria et al. (2007) allowed to estimate the sensitivity of the simulated streamflows to the parameters that control the surface and subsurface runoff generation, and the global-scale study of Chaney et al. (2015) evaluated the efficiency of the VIC model for monitoring global floods and droughts under a parameter uncertainty framework. The sensitivity to land use changes and the vegetation parameters associated to the different vegetation classes (i.e. Leaf Area Index and albedo) have also been explored and expressed for the different components of the water and energy balances from the VIC model (VanShaar et al., 2002; Chawla and Mujumdar, 2015; Bennett et al., 2018).

This work aims to contribute to the knowledge of the VIC model in a macroscale application over the headwater subwatersheds of an important basin located in the north of the Iberian Peninsula, the Duero River Basin. For this end, the hydrologic modeling exercise here developed has been divided into three interrelated parts:

- The calibration and validation of the VIC model for the selected subwatersheds of the study area.

- An integrated sensitivity analysis for all the subwatersheds focused on the soil parameters chosen for the calibration.

- A final assessment of equifinality and the efficiency of the calibration algorithm that links the calibration and sensitivity analysis results.

The Duero River Basin has been investigated in various previous studies and the main issues addressed are: the temporal trend of water supply and its relation to precipitation, temperature and plant cover changes (Ceballos-Barbancho et al., 2008); the hydrologic response to land-cover changes (Morán-Tejeda et al., 2010, 2012a), the impacts of different climate oscillations (Morán-Tejeda et al., 2010) and its response to the North Atlantic Oscillation (Morán-Tejeda et al., 2011a); the characteristics of the different existing river regimes (Morán-Tejeda et al., 2011b) and the effects of reservoirs on them (Morán-Tejeda et al., 2012b). However, all these studies were based on statistical analyses of different hydroclimatic and land-surface variables and contributed to a better understanding of the hydrologic behavior of the Duero River Basin. Therefore, the hydrologic modeling analysis carried out in this work can provide an added value to this set of issues since the potentialities of a macroscale hydrologic model such as the VIC model have been examined in detail for this river.

In Sect. 2 and 3 the study area and the methods are described. Sect. 4 gathers the results of the three-part approach and Sect. 5 corresponds to the discussion of the key results. Finally, the main conclusions of this study are provided in Sect. 6.

## 2. Study area

The Duero River Basin constitutes the largest basin of the Iberian Peninsula with a surface of 98 073 km$^2$. It is a shared territory between Spain and Portugal, characterized by a high water contribution (~ 15 000 hm$^3$/year). The study is focused on the Spanish part of the basin (Fig. 1), which represents the 80% of the area (78 859 km$^2$). Most of this territory constitutes a plain surrounded by mountainous chains, thus configuring two topographic areas well differentiated. The large depression is filled with sediments of the Tertiary and the Quaternary, constituting a complex hydrogeologic environment. The lithology of the northern mountains consists of siliceous, calcareous and carbonated rocks with local small aquifers to the west part and aquifers of greater capacity to the east. The south system harbors rocks of low permeability and is dominated by a granite batholith. Finally, the eastern mountainous areas hold a silicic core enclosed by carbonated rocks with a high presence of karstic aquifers.

The basin presents a predominant Mediterranean climate with a mean annual precipitation volume of 50 000 hm$^3$ which is mostly lost into the atmosphere through evaporative fluxes (~ 35 000 hm$^3$/year). Most precipitation is concentrated in the mountainous areas reaching values above 1500 mm/year to the north of the basin and values slightly below 1000 mm/year to the south and east. As for the most part of the Iberian Peninsula, precipitation exhibits a very irregular intra-annual distribution, being concentrated in spring and fall and almost nonexistent during summer. Winter months are cold with a mean temperature of 2ºC in January, while summer is soft with maximum temperatures occurring in July (~ 20.5ºC).

The Duero River Basin is regulated by a total of 31 reservoirs where the streamflow records are estimated through a water balance of the daily storages and water releases. The streamflow monitoring is also carried out in a large network of ca. 200 gauging stations where the streamflow records are calculated through the rating curves.

## 3. Methods

### 3.1 Hydrologic dataset

The streamflow records were gathered in a monthly basis from the Spanish Centre for Public Work Experimentation and Study (CEDEX, *Centro de Estudios y Experimentación de Obras Públicas*) database for all the reservoirs and gauging stations of the Duero Basin. An analysis of the percentage of gaps in the time series revealed that the period from October 2000 to September 2011 presents less than 5% of missing values for all the time series. Therefore, it was chosen as the study period for this work. A reference hydrologic network was then defined applying the criterion of the absence of upstream hydraulic structures (Whitfield et al., 2012). This selection reduced the number of reservoirs and gauging stations for the analysis to 16 reservoirs and 15 gauging stations covering the headwaters of the Duero River Basin (Fig. 1). Table 1 collects the main characteristics of these subwatersheds: area ($km^2$), mean elevation (m), averaged annual precipitation ($P_{an}$, mm/year), potential evapotranspiration ($PET_{an}$, mm/year), and streamflow ($Q_{an}$, $hm^3$/year), for the study period.

Precipitation, maximum temperature and minimum temperature data were extracted from two high-resolution (~ 5 km x 5 km) daily gridded datasets: SPREAD (Serrano-Notivoli et al., 2017) for precipitation data and STEAD (available at http://dx.doi.org/10.20350/digitalCSIC/8622) for temperature data. Both datasets cover Peninsular Spain and the Balearic and Canary Islands and were built with information from observed precipitation and maximum and minimum temperature at a varying number of meteorological stations provided by several administrations including the Spanish Meteorological Agency (AEMET) and some hydrologic confederations. Atmospheric pressure, incoming shortwave radiation, incoming longwave radiation, vapor pressure and wind speed data were taken from daily outputs of high resolution (0.088º, ~10 km) simulations carried out with the Weather Research and Forecasting (WRF) model driven by the ERA-Interim Reanalysis data for the spatial domain of the Iberian Peninsula (García-Valdecasas Ojeda et al., 2017).

### 3.2 Hydrologic modeling

3.2.1 The VIC model

The VIC model (Liang et al., 1994, 1996) is a semi-distributed macroscale hydrologic model that computes both the water and the energy balance within the grid cell. The sub-grid variability in land cover classes is evaluated statistically, and a spatially heterogeneous structure for the infiltration capacity is assumed using the formulation described in the Xinanjiang model (Zhao et al., 1980). This approach takes into account the sub-grid variability in the soil moisture storage capacity.

The water balance in the VIC model considers three types of evaporation: evaporation from bare soil, evaporation from the canopy layer for each vegetation class and transpiration from the different types of vegetation. Potential evapotranspiration is calculated from the Penman-Monteith equation, and it represents the atmospheric demand for water vapor. Actual evapotranspiration in a grid cell is obtained as the sum of the three evaporation types weighted by the fraction of the area corresponding to each land cover class.

Different algorithms for the runoff generation process must be contemplated depending on the number of soil layers defined. The three-layer VIC model (VIC-3L) is the most common application and has been chosen for this work since it is a modification of the two-layer VIC model (VIC-2L) originally developed to better represent the runoff generation process (Liang et al., 1996). Surface runoff is generated through an infiltration excess applying the Xinanjiang formulation (Zhao et al., 1980) to the upper two soil layers:

$$Q_d = \begin{cases} P - z_2 \cdot (\theta_S - \theta_2) + z_2 \cdot \theta_S \cdot \left(1 - \dfrac{i_0 + P}{i_m}\right)^{1+b_i}, & P + i_0 \le i_m \\ P - z_2 \cdot (\theta_S - \theta_2), & P + i_0 \ge i_m \end{cases} \qquad (1)$$

For each time step $Q_d$ [L] is the surface (direct) runoff, $P$ [L] is the precipitation, $z_2$ [L] is the depth of the upper two soil layers, $\theta_2$ is their volumetric soil moisture content, $\theta_S$ is their porosity, $i_m$ [L] is the maximum infiltration capacity, $i_0$ [L] is the infiltration capacity that corresponds to the soil moisture at that time step and $b_i$ is the infiltration shape parameter.

Baseflow is generated in the third soil layer following the Arno formulation (Franchini and Pacciani, 1991), and is expressed as:

$$Q_b = \begin{cases} \dfrac{D_S \cdot D_m}{W_S \cdot \theta_S} \cdot \theta_3, & 0 \le \theta_3 \le W_S \cdot \theta_S \\ \dfrac{D_S \cdot D_m}{W_S \cdot \theta_S} \cdot \theta_3 + \left(D_m - \dfrac{D_S \cdot D_m}{W_S}\right) \cdot \left(\dfrac{\theta_3 - W_S \cdot \theta_S}{\theta_S - W_S \cdot \theta_S}\right)^2, & \theta_3 \ge W_S \cdot \theta_S \end{cases} \qquad (2)$$

Here, $Q_b$ [L] is the baseflow for each time step, $D_m$ [L] is the maximum baseflow, $D_S$ is a fraction of $D_m$, $\theta_3$ is the volumetric soil moisture content of the soil layer 3, $\theta_S$ is the porosity in this layer and $W_S$ is a fraction of $\theta_S$. The baseflow recession curve is divided into two parts: a linear part for lower values of $\theta_3$ and a non-linear (quadratic) part for higher values of $\theta_3$.

The characteristic large grid size for macroscale models makes that the VIC snow model conceptualizes the snow processes partitioning each grid cell of the spatial domain into snow bands, thereby accounting for the sub-grid variability in topography, land uses and precipitation. The snow model is applied separately to each snow band and land class and the outputs consist of the snow depth and the snow water equivalent for each grid cell. The snowpack is represented as a two-layer model that solves the energy and the mass balance and determines whether the snowpack is subject to accumulation or ablation, making the model suitable for applications in any part in the world (Liang et al., 1994, 1996).

### 3.2.2 Soil and vegetation parameters

The required soil parameters for the application of the VIC model were obtained from SoilGrids1km (Hengl et al., 2014) and EU-SoilHydroGrids ver1.0 (Tóth et al., 2017), all of them with a spatial resolution of 1 km x 1 km. In both datasets the different soil properties are provided for seven soil depths up to 2 m (0, 5, 15, 30, 60, 100 and 200 cm). These soil parameters are: (1) bulk density and soil textural classes of the United States Department of Agriculture (USDA) from SoilGrids1km; and (2), field capacity, saturated hydraulic conductivity, porosity and wilting point from EU-SoilHydroGrids ver1.0.

The VIC model handles land uses information as a set of vegetation parameters for the different vegetation classes specified in a vegetation library. Here the UMD Global land cover classification (Hansen et al., 2000) was chosen with a spatial resolution of 1 km x 1 km, and the vegetation parameters (i.e. Leaf-Area Index, roots depth and roots coverage) were fixed for each vegetation class following the recommendations of the GLDAS Project (https://ldas.gsfc.nasa.gov/gldas/vegetation-parameters).

### 3.2.3 Routing procedure

The water balance mode of the VIC model allows to simulate the surface runoff and the baseflow for each grid cell of the spatial domain where the model is implemented. Since the conceptualization of the VIC model does not include the horizontal transport processes between contiguous grid cells, the runoff generated needs to be routed to a certain outlet in order to determine simulated values of streamflow that can be compared with existing observations. An approach consisting of a monthly aggregation of the total runoff (the sum of surface runoff and baseflow) for all the grid cells within a given subwatershed, weighted by the fractional area of each cell inside the subwatershed, was selected for this work. This approach has the advantage of working with the real boundaries of the subwatersheds and is not dependent on the resolution of the grid cells.

### 3.2.4 Model implementation

The VIC model was implemented in the water balance mode at a daily time step and with a spatial resolution of 0.05º (~ 5 km x 5 km) for the 31 studied subwatersheds (Fig. 2). The meteorological forcings were interpolated to the grid cells of the spatial domain following a nearest neighbor assignment. The soil parameters and the elevation were averaged for each grid cell, and the vegetation parameters were kept at the original resolution of 1 km x 1 km because the VIC model allows to consider the sub-grid variability of the land uses for each grid cell. As in the soil database, the depth of each grid cell was set at 2 m, fixing the thickness of the first soil layer ($d_1$) at 0.1 m and varying the thicknesses of the second ($d_2$) and third ($d_3$) layers during the calibration and the sensitivity analysis.

The outputs of the VIC model were finally aggregated into monthly values for each subwatershed in order to accomplish the calibration, validation and sensitivity analysis.

### 3.3 Calibration and Validation

The model was calibrated for the period from October 2000 to September 2009 choosing the Nash-Sutcliffe Efficiency (*NSE*, Nash and Sutcliffe, 1970) as the objective function. The *NSE* was calculated by comparing the monthly observations of streamflow with the monthly aggregated total runoff simulated by the VIC model. Table 2 indicates the selected parameters for the calibration process together with their upper and lower bounds. The calibration was carried out using the Shuffled-Complex-Evolution Algorithm (SCE-UA) of Duan et al. (1994). A spin-up period of ten years, previous to the calibration period, was simulated in order to ensure that the soil moisture content of the three soil layers reached an equilibrium, and therefore the initial conditions did not affect the calibration process. The model was then validated for the period October 2009 – September 2011. The goodness-of-fit for both calibration and validation exercises was evaluated through the *NSE* and the coefficient of determination $r^2$.

### 3.4 Sensitivity Analysis

The Standardized Regression Coefficients (SRC) method (Saltelli et al., 2008) aims to study the propagation of uncertainty from model inputs to outputs. The SRC method is focused on the behavior of the model outputs in relation to a certain set of parameters once the boundary conditions (i.e. the meteorological forcings) and the initial conditions (i.e. the soil moisture content of the three soil layers) have been fixed. This sensitivity analysis method requires two elements: first, a Monte Carlo simulation where the model is run with a specified number of parameter samples; and second, a multiple linear regression of each model output of interest as a linear function of the parameters.

The sensitivity analysis was carried out for each subwatershed considered in the study area. As in the calibration process, the period October 2000 – September 2009 was chosen and ten years of spin-up prior to the study period were run.

3.4.1 Monte Carlo simulation

A parametric space is defined through the selection of several parameters and their upper and lower bounds. Here, a 5-dimensional parametric space was established choosing the five calibration parameters and considering the upper a lower bounds specified in Table 2. A sampling method is then applied to the parametric space, extracting a large-enough sample of parameters for the Monte Carlo simulation. The Latin Hypercube Sampling (LHS) method (Iman and Conover, 1982) was applied for this step extracting a total of 10 000 random samples. This process allowed to define a sampling matrix, $\mathbf{\Theta}$, of order $m$ x $n$, where $m$ represents the number of samples ($m$ = 10 000) and $n$ the number of parameters for the analysis ($n$ = 5). The model was finally run for each parameter combination (i.e. row) of $\mathbf{\Theta}$.

The Monte Carlo simulation was also used for assessing equifinality and the efficiency of the calibration algorithm by studying the response given by each parameter sample in terms of the *NSE*. The results were compared with the *NSE* determined during the calibration period.

3.4.2 Multiple linear regression

The outputs of interest from the VIC model were those components included in the water balance: surface runoff ($Q_d$), baseflow ($Q_b$), total runoff ($Q_t$), actual evapotranspiration (*AET*) and the soil moisture content of the three soil layers ($SM_1$, $SM_2$ and $SM_3$). For each component and for each run of the Monte Carlo simulation, the mean value of the simulated series was calculated, and a multiple linear regression model was then adjusted relating the mean values of each component with the sampling parameters:

$$\boldsymbol{y} = a_0 \cdot \begin{bmatrix} 1 \\ \vdots \\ 1 \end{bmatrix} + \sum_{i=1}^{n} a_i \cdot \boldsymbol{\Theta}_i \qquad (3)$$

Where $\boldsymbol{y}$ is a column vector with the $m$ mean values of the component, $a_0$ is the intercept of the hyperplane, $a_i$ is the regression coefficient of the parameter $i$ and $\boldsymbol{\Theta}_i$ is the column of the sampling matrix corresponding to the parameter $i$. The standardized regression coefficients, $\beta_i$, are then calculated for each parameter:

$$\beta_i = \frac{\sigma_{\boldsymbol{\Theta}_i}}{\sigma_{\boldsymbol{y}_p}} \cdot a_i \qquad (4)$$

Here, $\sigma_{\boldsymbol{\Theta}_i}$ and $\sigma_{\boldsymbol{y}_p}$ are the standard deviations of $\boldsymbol{\Theta}_i$ and the predicted values of $\boldsymbol{y}$, respectively. $\beta_i^2$ represents the relative contribution of the parameter $i$ to the variance of the model output of interest, being $\sum_{i=1}^{n}(\beta_i^2) \leq 1$ and equal to the coefficient of determination $r^2$ of the adjustment. A threshold of $r^2 \geq 0.7$ is usually defined for assuming that the fitted model has a good linear behavior, and therefore the coefficients $\beta_i$ are valid measures of the sensitivity (Saltelli et al., 2006), although they can be robust and reliable measures even for nonlinear models (Saltelli et al., 2008). $\beta_i$ can take values between -1 and 1. A high absolute value of $\beta_i$ implies that the component is sensitive to the parameter and its sign indicates whether the effect is positive or negative.

## 4. Results

### 4.1 Calibration and validation

The values of *NSE* and $r^2$ for the calibration and validation periods are shown in Table 3. Figure 3 depicts the simulated streamflows during both periods together with the observed streamflows for six selected subwatersheds (R-2011, R-2037, R-2038, GS-3005, GS-3089 and GS-3150) located in different parts of the basin. The *NSE* for the calibration period presents values above 0.75 in 19 out of the 31 subwatersheds and reaches values above 0.85 in 10 subwatersheds, and the corresponding $r^2$ values are high too. For the validation period, both *NSE* and $r^2$ values are predominantly high, and generally lower than the corresponding ones for the calibration process, although minimum values of *NSE* below 0 were attained for 3 subwatersheds. Note that the results of the calibration and the validation processes are slightly better for the reservoirs, what could indicate a quality difference between the streamflow databases from the reservoirs and from gauging stations. Some of the high $r^2$ values are obtained for low *NSE* estimations, which indicates that the model is able to capture the intra-annual variability of the streamflow observations but is not able to reach a good fit for the peaks of streamflow (Table 3, Fig. 3). It is interesting to note that high *NSE* values were obtained for subwatersheds with varying sizes, with good fits for both small-sized (e.g. R-2011 and GS-3089) and medium-sized (e.g. R-2038 and GS-3005) subwatersheds, emphasizing the ability of the VIC model to provide accurate predictions of the streamflow across different spatial scales.

### *4.2 Integrated sensitivity analysis*

Through the application of the SRC method the $\beta$ coefficients for the five calibration parameters of the water balance components in the VIC model were obtained, and the results are shown in Fig. 4 (a to g) for all the subwatersheds. The $r^2$ value obtained from the multiple linear regression and the estimation of $r^2$ as the sum of the squares of $\beta$ coefficients are also depicted in Fig. 4 (h, i), reflecting very similar values. The results of the sensitivity analysis for the selected components to the parameters are given below in a component-by-component basis providing the necessary explanations when there is a strong dependency between them:

- $Q_d$: the values of $r^2$ are above 0.7 for all the subwatersheds (Fig. 4), fulfilling the criterion of enough linearity for interpreting the results of the sensitivity analysis. The strongest positive effect corresponds to the parameter $b_i$, which means that a higher value of $b_i$ leads to more surface runoff. This is clearly evidenced in Eq. 1, where a relation of exponential type between $Q_d$ and $b_i$ is established. $D_m$ produces a negative effect on the surface runoff, suggesting that an increase of the maximum baseflow brings a reduction of the surface component under the assumption of the same meteorological forcings. $d_2$ also yields a negative effect on $Q_d$, and this effect is related to an increase in *AET*.

- $Q_b$: the values of $r^2$ mostly range from 0.5 to 0.7, with some values close to 0.8 (Fig. 4). In this case the linearity criterion is hardly reached and therefore it is difficult to interpret the $\beta$ coefficients. However, it is interesting to note that, with the exception of $d_2$, the $\beta$ coefficients of the parameters are characterized by a low dispersion. This is an indicator of the robustness of the VIC model response, and although the threshold of linearity is not always achieved, the dependency of $Q_b$ with respect to these parameters can be accepted. As expected from the previous analysis of the surface component of the runoff, $b_i$ reflects a strong negative effect. The positive effects now correspond to $D_S$ and $D_m$. This is obvious in the case of $D_m$ but not so evident for $D_S$ since a higher value of $D_S$ only means that the baseflow law tends to be more linear (see Eq. 2). The amplitude of the $\beta$ coefficients for $d_2$ is broader than for the rest of the parameters but always negative except for two subwatersheds.

- $Q_t$: the total runoff exhibits an additive effect of the previous components for both $r^2$ and the $\beta$ coefficients as it is computed through the sum of the surface runoff and the baseflow (Fig. 4). Thus, higher values of $b_i$, $D_S$ and $D_m$ lead to an increase of $Q_t$ and higher values of $d_2$ produce a negative effect on $Q_t$ due to a rise of *AET*. This component is of particular interest given that it is the component subject to calibration in this work. In order to provide a better understanding on its behavior the spaghetti plots of the Monte Carlo simulation for an example subwatershed (R-2038) are depicted in Fig. 5. As shown there, the time series of the observed streamflow (Fig. 5c) falls into the range of responses of the model for almost all the study period and therefore one or more sets of parameters will afford a god fit with the observations.

- *AET*: this component and $Q_t$ are linked through the law of conservation of mass applied to the system defined by each subwatershed, being the precipitation equal to the sum of $Q_t$, *AET* and the variation of the storage in the hydrologic system. Moreover, the study period for the sensitivity analysis is long-enough to neglect the last term of the water balance equation, and the precipitation is fixed for each subwatershed as a boundary condition. In consequence, the linearity of both $Q_t$ and *AET* with respect to the parameters must be similar and the $\beta$ coefficients for *AET* are essentially identical to the corresponding ones for $Q_t$ but with opposite signs (Fig. 4). Figure 5d shows the spaghetti plots of this component together with the potential evapotranspiration (*PET*) profile. The reason of the existence of some values of *AET* above the *PET* curve responds to the internal handling of the Penman-Monteith equation used in the VIC model because various different approaches are considered when computing the potential evapotranspiration, and the curve presented in Fig. 5d corresponds to the current vegetation parameters.

- $SM_1$: the values of $r^2$ are widely scattered and range from 0.35 to values above 0.9 (Fig. 4). The nature of such a scattered distribution may be an outcome of the closeness between the soil moisture profiles in this layer, making it difficult to adjust a multiple linear regression model to its mean values. Similarly to the case of $Q_b$, most of the $\beta$ coefficients present a relatively low dispersion and subsequently the results of the sensitivity analysis can be interpreted. The negative effects mainly concern to $b_i$ and $D_m$, demonstrating that a higher exponent in the surface runoff equation and a higher maximum baseflow are related to lower soil moisture values for the upper soil layer. On the other hand, the positive effects are associated with increasing values of $d_2$ despite of revealing highly dispersed $\beta$ coefficients. The spaghetti plots of $SM_1$ (Fig. 5e) seem to reproduce the $PET$ cycles and this results from the evaporative fluxes themselves as the transpiration process occurs from the roots of the vegetation.

- $SM_2$: it is the component with the highest linearity with regard to the calibration parameters, proffering values of $r^2$ very close to 1 for all the subwatersheds (Fig. 4). In this case $d_2$ dominates the sensitivity of $SM_2$ with a noticeable positive effect (i.e. values of $\beta$ near to 1). The $PET$ cycles are also markedly represented in the soil moisture profiles of this layer (Fig. 5f).

- $SM_3$: even though the values of $r^2$ lie between 0.6 and 0.7 predominantly, some of them fall below 0.6 with minimum values close to 0.4 (Fig. 4). Once again the dispersion of the $\beta$ coefficients is relatively low and in this occasion this is also true for $d_2$. Baseflow takes place from this layer and this is reflected in the $\beta$ coefficients corresponding to $D_S$, $W_S$ and $D_m$, which present opposite signs and similar absolute values to the calculated ones for $Q_b$. As for the previous soil layers, the $PET$ cycles are present too but here there is a lag in the valleys of the soil moisture profiles due to the delay in the baseflow generation process (Fig. 5g).

### 4.3 Assessing equifinality and the efficiency of the calibration algorithm

Equifinality and the efficiency of the calibration algorithm were assessed through the evaluation of the *NSE* values for the Monte Carlo simulations of all the subwatersheds by comparing the total runoff of each simulation with the observed streamflow during the calibration period. For this purpose, two counts of the number of simulations satisfying certain criteria were carried out: first, the number of simulations for each subwatershed presenting *NSE* values above the *NSE* determined during the calibration ($NSE_{cal}$) minus 0.05 was used as indicator of equifinality of the VIC model and the parameter samples; and second, the number of simulations with *NSE* values above $NSE_{cal}$ hinted at the efficiency of the SCE-UA algorithm in finding the optimal set of parameters producing the best fit with the streamflow records. The results of this exercise are expressed in Table 4.

It is clear that for the majority of the subwatersheds there are many simulations with *NSE* values very close to the optimal model, and in some cases the number of simulations is fairly high (> 3000). This can be also appreciated when the columns of the sampling matrix are plotted against the *NSE* of each simulation in a "dotty plot". Figure 6 shows the dotty plots of the five parameters of the calibration for two subwatersheds (R-2038 and GS-3089) as an example of this analysis. For both subwatersheds the *NSE* values of the simulations are above 0 and the points clouds are concentrated on the top of the diagrams, suggesting that a high number of them are close to the optimal fit (see also Table 4). The shape of the dotty plots supplies useful information about the behavior of the parameter samples as a set. For example, the dotty plot of $d_2$ for the subwatershed R-2038 reflects a trend to produce high *NSE* values when the parameter values are near to the upper bound. The optimum was reached for $d_2 = 0.8995$ m, while the rest of the fitted parameters were located between the fixed limits. Also, most of the *NSE* values were below $NSE_{cal}$ when the second count was executed, implying that the SCE-UA algorithm was highly efficient in searching for the optimal set of parameters.

## 5. Discussion

The results of the calibration and validation processes are comparable to other studies using hydrologic models developed in northern Spain (Morán-Tejeda et al., 2014) and recently in the south of Spain (Pellicer-Martínez and Martínez-Paz, 2018; Yeste et al., 2018). It is to be expected that the application of a single model structure over an heterogeneous spatial domain, such as the Duero River Basin, does not conduct to a good adjustment of the simulated streamflow with the observations for all the studied subwatersheds. Furthermore, the existence of other potential pressures over the water resources may be responsible for those cases where the calibration and/or validation showed poor results, and therefore further research is required in order to identify the origin of the biases with respect to the observations of the simulated streamflows for these subwatersheds. Nevertheless, the results of the calibration and the validation suggest that the macroscale application of the VIC model carried out in this study performs well for a large number of subwatersheds in the Duero River Basin. Moreover, the routing procedure has proven to be accurate and efficient in this work, permitting its application in other studies using hydrologic models that operate over the grid cell in a similar mode to the VIC model.

Concerning the sensitivity analysis, the application of the SRC method allowed a deep understanding of the existing relationships between the components of the water balance in the VIC model and the selected parameters for the calibration as long as the linearity criterion was fulfilled. Even when the coefficient of determination of the fitted model did not satisfy the linearity criterion, the relatively low dispersion of the $\beta$ coefficients permitted the interpretation of the results. Special attention deserves the component $Q_t$ since it is the component that was compared with the streamflow observations during the calibration and validation processes. The sensitivity of this component to the five soil parameters reflected an additive effect of the sensitivity measures of $Q_d$ and $Q_b$ as $Q_t$ is calculated as the sum of the surface and the subsurface components. $Q_t$ was mainly sensitive to $b_i$ and $d_2$, and this is consistent with the sensitivity measures for the simulated streamflow carried out in Demaria et al. (2007) for four studied subwatersheds.

At the sight of the results of the equifinality assessment, it is unavoidable accepting that no parameter set leads to a single optimal model, or in other words, that there are many parameter samples with performances as good as the optimum calculated with the calibration algorithm. As in the GLUE method (Beven and Binley, 1991; Beven, 2012), this fact could be the starting point of the calibration process, in which a measure of belief is associated to each parameter set according to the degree of proximity to the optimum. This will be an interesting research line for further investigation in the Duero River Basin. In any case, we consider that the use of a calibration algorithm provides a first-look into the goodness-of-fit response surface of the hydrologic model in a computationally more efficient way than the Monte Carlo experiment, serving as a sign of the goodness-of-fit of the overall parameter samples.

## 6. Conclusions

The main conclusions of this work can be summarized as follows:

[1] The calibration and validation of the VIC model reflected good results for most of the studied subwatersheds in the Duero River Basin with a predominance of high *NSE* values. The results were slightly better for the reservoirs than for the gauging

stations and this may be a consequence of a quality difference between the streamflow databases. The calibration and validation processes showed poor results in a few subwatersheds. This may be caused by the existence of pressures over the water resources that have not been taken into account in the modeling exercise. However, this is out of the scope of this work since the main interest is placed on the macroscale application of the VIC model, which has shown to perform well for a great part of the Duero River Basin.

[2] The $\beta$ coefficients calculated during the sensitivity analysis allowed to quantify the sensitivity of the water balance components to the selected parameters for the calibration. The surface runoff and the soil moisture content of the soil layer 2 were the components with the highest linearity and were mainly dominated by the values of the infiltration shape parameter and the thickness of soil layer 2, respectively, both with a positive effect. The total runoff presented a combined behavior from the surface runoff and the baseflow components, and the sensitivity analysis yielded similar results to other sensitivity measures previously reported in the literature. The potential evapotranspiration cycles were noticeable in the whole soil profile and more evidently in the upper two soil layers.

[3] A final exercise for assessing equifinality and the efficiency of the calibration algorithm was carried out, finding that there are many parameter sets with *NSE* values as high as the *NSE* determined during the calibration. The calibration algorithm was efficient and reached the optimal fit for almost all the studied subwatersheds. The use of a calibration algorithm is also in line with other possible practical applications of the VIC model for studying the impacts of climate change on water resources in the Duero River Basin, where a parameter set must be chosen prior to the simulations using climate change data.

**Acknowledgements**

All the simulations were conducted in the ALHAMBRA cluster (http://alhambra.ugr.es/) of the University of Granada. This work was partially funded by the Spanish Ministry of Economy and Competitiveness projects CGL2013-48539-R and CGL2017-89836-R, with additional support from the European Community Funds (FEDER). The first author was supported by the Ministry of Education, Culture and Sport of Spain (FPU grant FPU17/02098).

**Figure captions**

**Figure 1.** Duero River Basin and the 31 studied subwatersheds. The prefix "R-" denotes "Reservoir" and the prefix "GS-" denotes "Gauging Station".

**Figure 2.** VIC model implementation.

**Figure 3.** Time series of the observed streamflows along with the simulated ones for the calibration and de validation periods for six example subwatersheds.

**Figure 4.** (a to g) $\beta$ coefficients for the 31 subwatersheds. h) $r^2$ value from the multiple linear regression. i) $r^2$ estimated from the $\beta$ coefficients.

**Figure 5.** Spaghetti plots of the water balance components resulting from the Monte Carlo simulation for the subwatershed R-2038.

**Figure 6.** Dotty plots for two subwatersheds: a) R-2038 and b) GS-3089. Red dot corresponds to the calibrated value for the corresponding parameter using the SCE-UA algorithm.

# DUERO RIVER BASIN

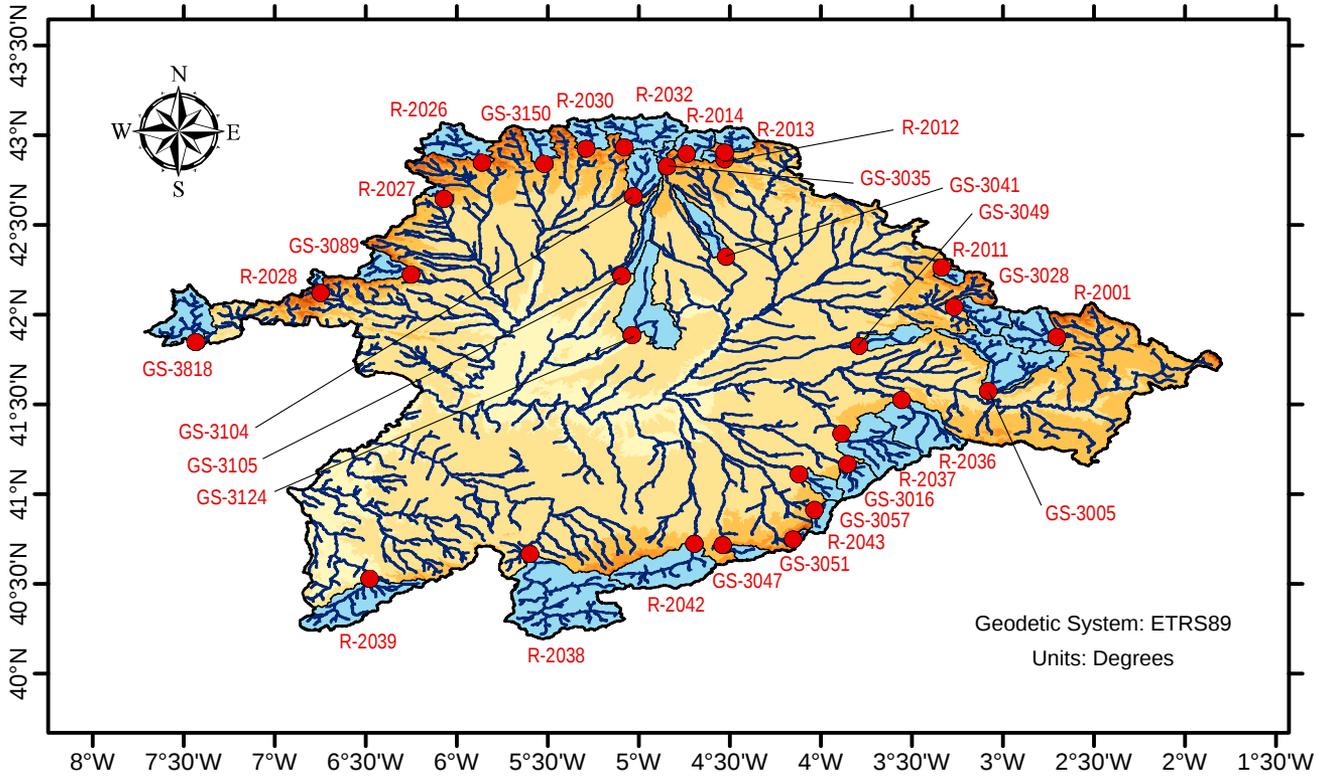

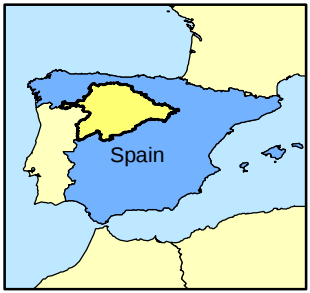

*SITUATION*

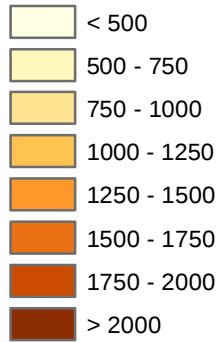

Elevation (m)

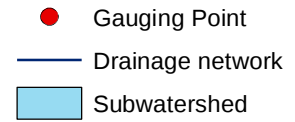

**Model resolution: 0.05º**

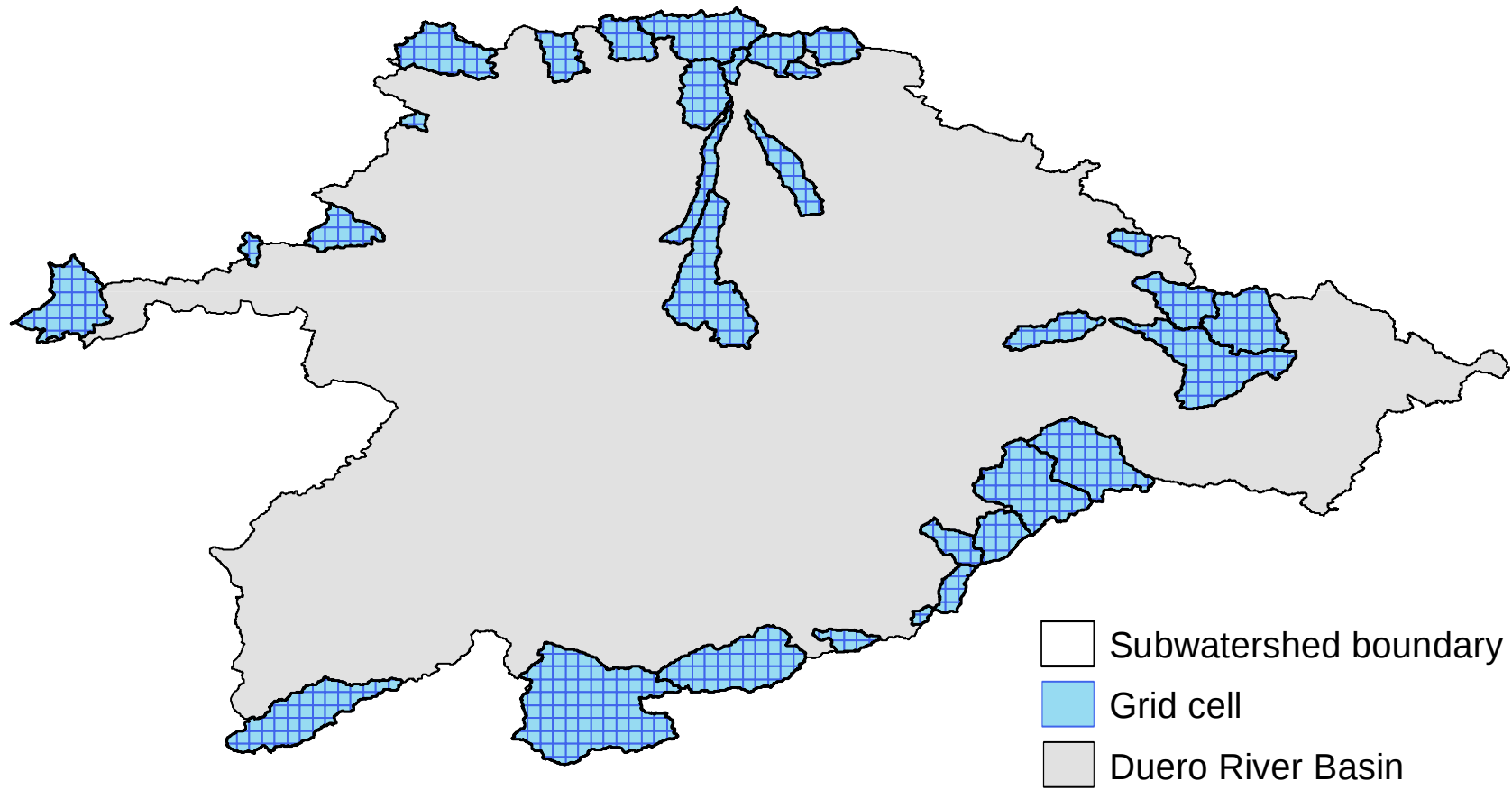

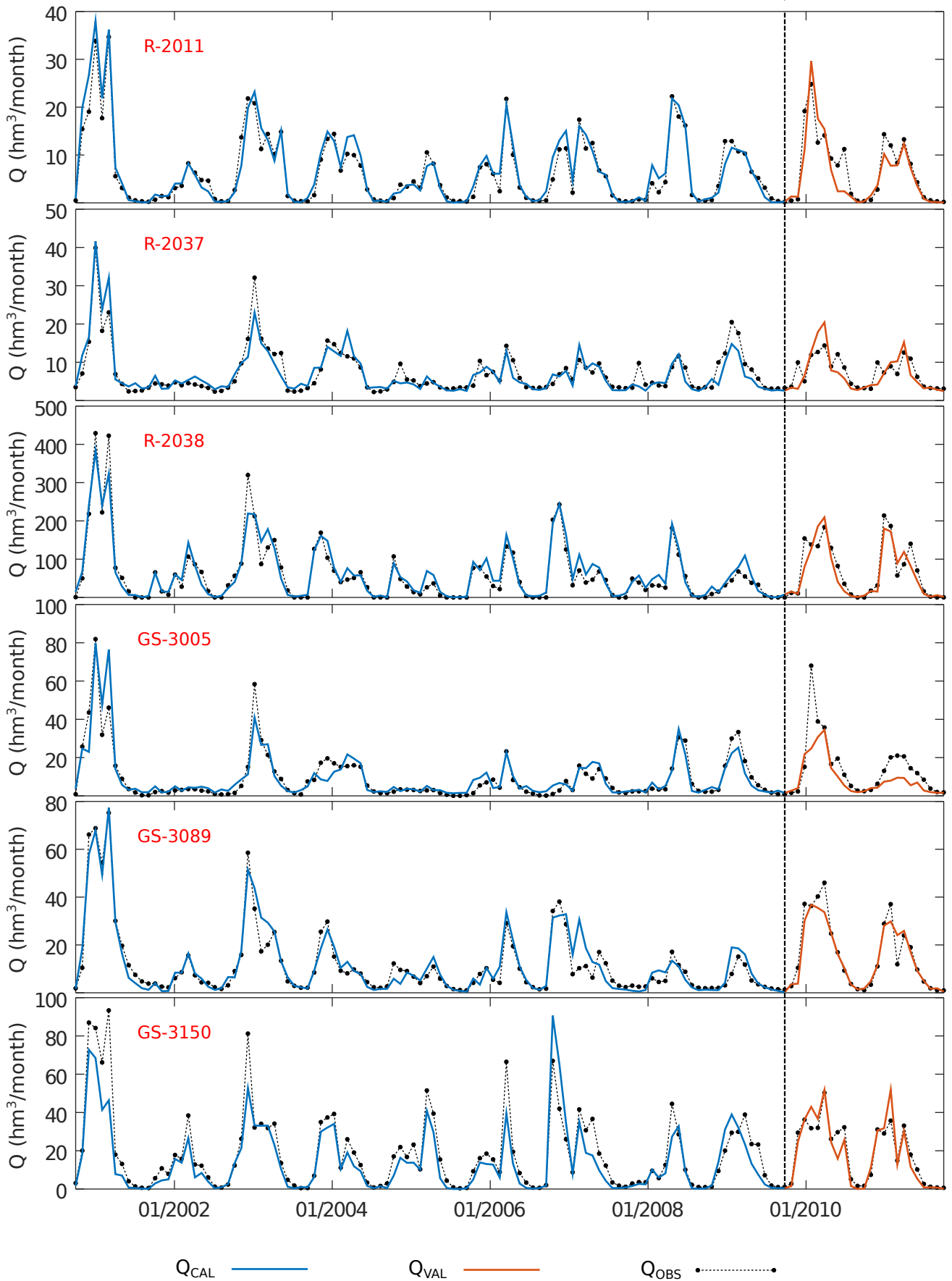

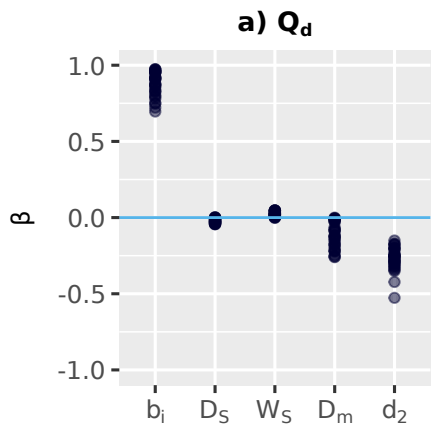

**a) $Q_d$**

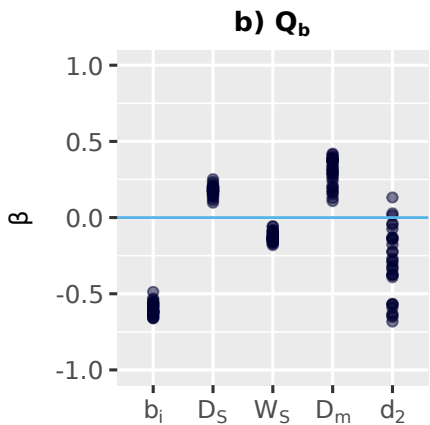

**b) $Q_b$**

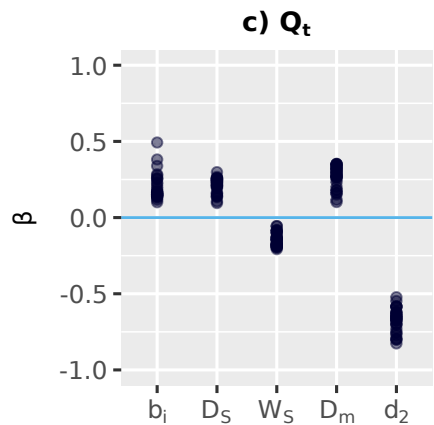

**c) $Q_t$**

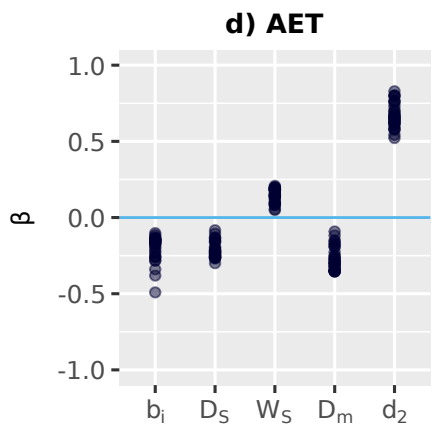

**d) AET**

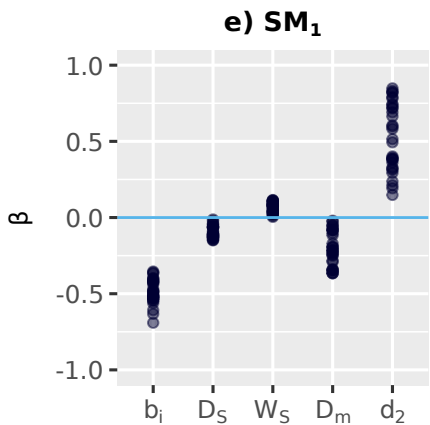

**e) $SM_1$**

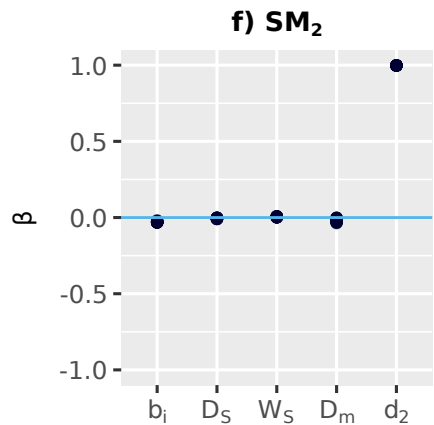

**f) $SM_2$**

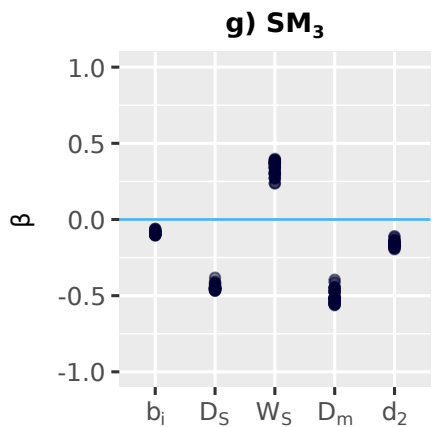

**g) $SM_3$**

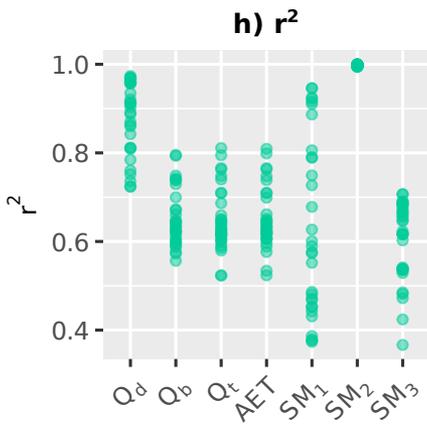

**h) $r^2$**

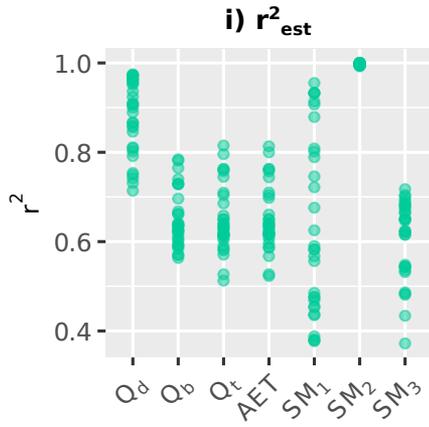

**i) $r^2_{est}$**

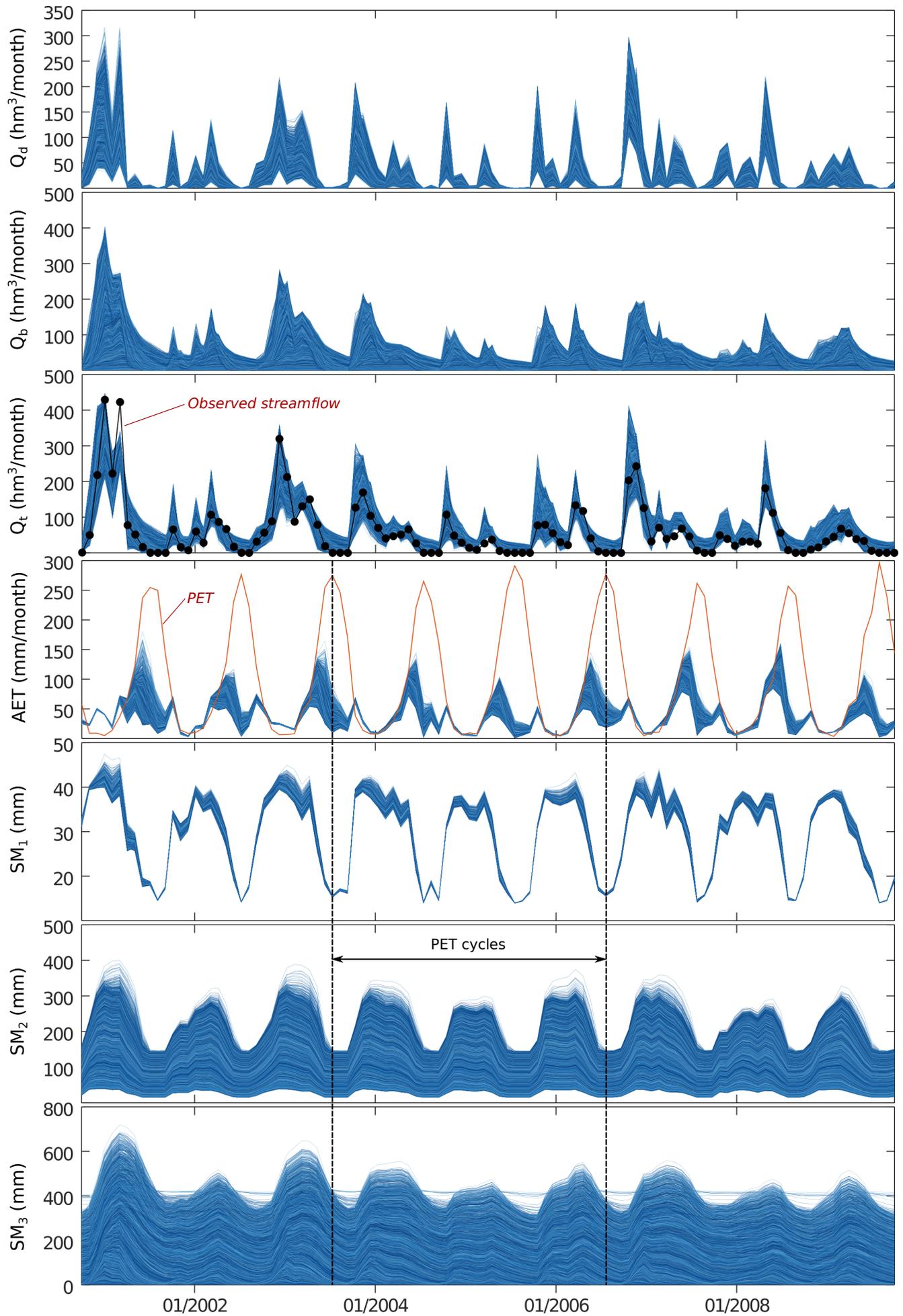

## a) R-2038

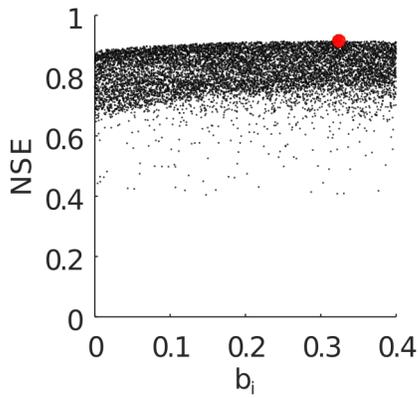
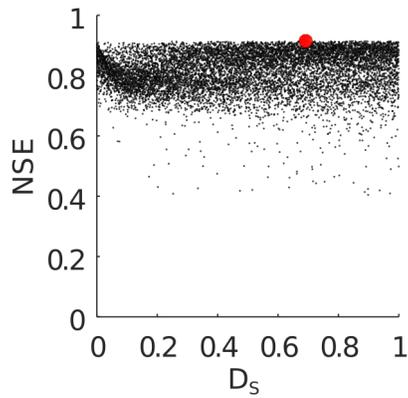
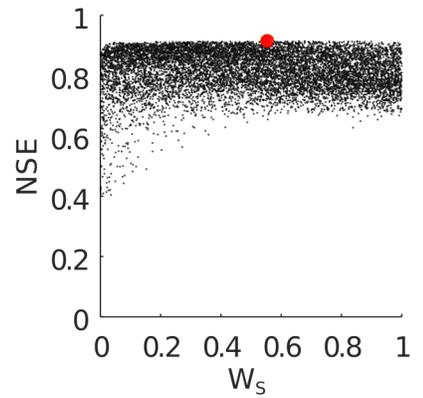

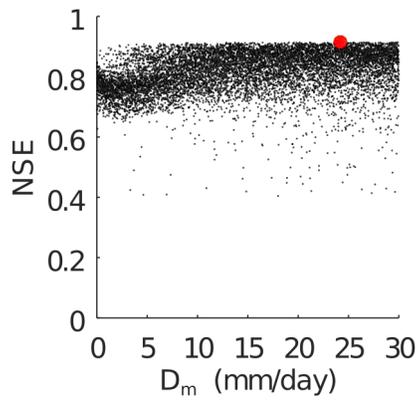
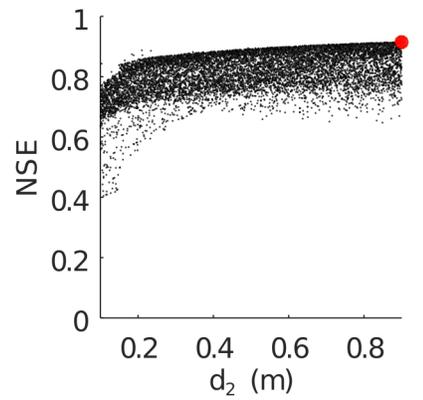

## b) GS-3089

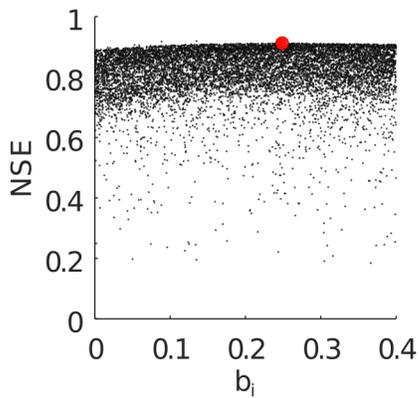
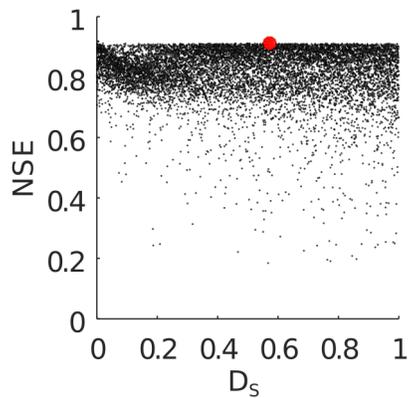
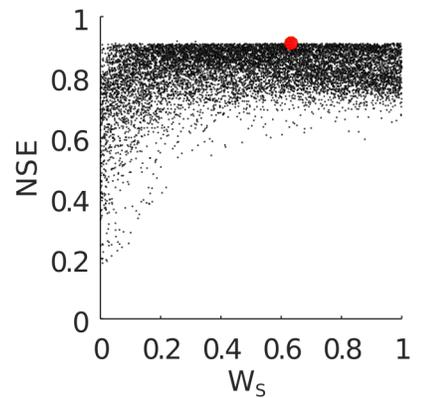

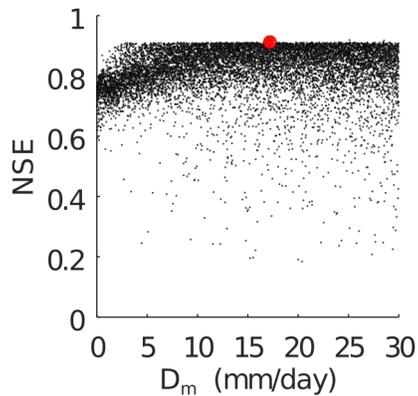
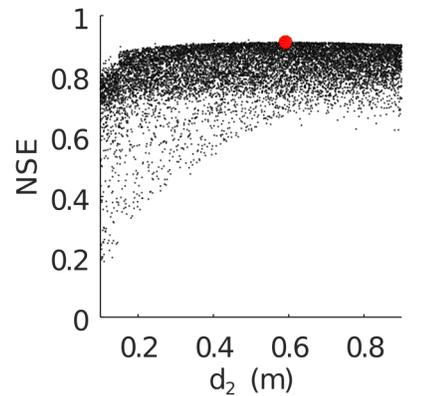

**Table 1.** Main characteristics of the 31 subwatersheds.

| Code | Name | Area (km$^2$) | Mean elevation (m) | $P_{an}$ (mm/year) | $PET_{an}$ (mm/year) | $Q_{an}$ (hm$^3$/year) |
|---|---|---|---|---|---|---|
| R-2001 | CUERDA DEL POZO | 546.7 | 1319 | 1122 | 900 | 188.7 |
| R-2011 | ARLANZON | 106.7 | 1440 | 1293 | 798 | 77.6 |
| R-2012 | CERVERA DE RUESGA | 54.3 | 1284 | 1005 | 859 | 86.1 |
| R-2013 | REQUEJADA, LA | 220.7 | 1378 | 1022 | 780 | 153.2 |
| R-2014 | CAMPORREDONDO | 229.6 | 1673 | 1457 | 747 | 223.9 |
| R-2026 | BARRIOS DE LUNA | 482.9 | 1496 | 1468 | 757 | 400.1 |
| R-2027 | VILLAMECA | 45.8 | 1180 | 946 | 941 | 34.1 |
| R-2028 | MONCABRIL (SISTEMA) | 62.9 | 1712 | 1242 | 744 | 93.5 |
| R-2030 | PORMA / JUAN BENET | 250.3 | 1412 | 1251 | 753 | 304.5 |
| R-2032 | RIAÑO | 592.3 | 1451 | 1569 | 748 | 623.7 |
| R-2036 | LINARES DEL ARROYO | 761.3 | 1111 | 515 | 1222 | 52.4 |
| R-2037 | BURGOMILLODO | 803.1 | 1097 | 573 | 1232 | 87.0 |
| R-2038 | SANTA TERESA | 1845.4 | 1326 | 882 | 1235 | 734.8 |
| R-2039 | AGUEDA | 788.4 | 895 | 1189 | 1333 | 392.7 |
| R-2042 | CASTRO DE LAS COGOTAS | 853.3 | 1279 | 527 | 1231 | 89.0 |
| R-2043 | PONTON ALTO | 150.4 | 1582 | 990 | 993 | 78.3 |
| GS-3005 | OSMA | 893.1 | 1090 | 728 | 1084 | 121.0 |
| GS-3016 | PAJARES DE PEDRAZA | 281.3 | 1298 | 634 | 1196 | 61.6 |
| GS-3028 | SALAS DE LOS INFANTES | 353.5 | 1257 | 988 | 926 | 104.7 |
| GS-3035 | OTERO DE GUARDO | 69.2 | 1492 | 1702 | 838 | 29.3 |
| GS-3041 | VILLALCAZAR DE SIRGA | 307.7 | 929 | 664 | 1011 | 35.5 |
| GS-3047 | MEDIANA DE VOLTOYA | 130.4 | 1347 | 518 | 1128 | 15.4 |
| GS-3049 | CABAÑES DE ESGUEVA | 270.2 | 995 | 658 | 1046 | 24.7 |
| GS-3051 | ESPINAR, EL | 36.7 | 1610 | 905 | 1032 | 18.8 |
| GS-3057 | VILLOVELA DE PIRON | 202.0 | 1183 | 615 | 1212 | 33.0 |
| GS-3089 | MORLA DE LA VALDERIA | 281.1 | 1369 | 1001 | 945 | 146.1 |
| GS-3104 | VILLAVERDE DE ARCAYOS | 371.1 | 1146 | 1041 | 942 | 135.3 |
| GS-3105 | SANTERVAS DE CAMPOS | 277.1 | 900 | 591 | 1107 | 28.8 |
| GS-3124 | MEDINA DE RIOSECO | 908.4 | 790 | 450 | 1315 | 37.2 |
| GS-3150 | PARDAVE | 223.8 | 1448 | 1356 | 787 | 220.8 |
| GS-3818 | RABAL | 597.9 | 678 | 1328 | 1146 | 305.7 |

**Table 2.** Selected parameters for the calibration.

| Parameter | Units | Lower bound | Upper bound | Description |
|---|---|---|---|---|
| $b_i$ | - | $10^{-5}$ | 0.4 | Infiltration shape parameter (see Eq. 1) |
| $D_S$ | - | $10^{-9}$ | 1 | Fraction of $D_m$ where non-linear baseflow starts (see Eq. 2) |
| $W_S$ | - | $10^{-9}$ | 1 | Fraction of the porosity of soil layer 3 where non-linear baseflow starts (see Eq. 2) |
| $D_m$ | mm/day | $10^{-9}$ | 30 | Maximum baseflow (see Eq. 2) |
| $d_2$ | m | 0.1 | 0.9 | Thickness of soil layer 2 |

**Table 3.** Values of *NSE* and $r^2$ for the calibration and validation periods.

| Code | $NSE_{cal}$ | $NSE_{val}$ | $r^2_{cal}$ | $r^2_{val}$ |
|------|------|------|------|------|
| R-2001 | 0.8520 | 0.4813 | 0.9494 | 0.8006 |
| R-2011 | 0.9237 | 0.7606 | 0.9696 | 0.8947 |
| R-2012 | 0.2723 | 0.3758 | 0.8983 | 0.8733 |
| R-2013 | 0.8263 | 0.8141 | 0.9263 | 0.9152 |
| R-2014 | 0.8777 | 0.8736 | 0.9370 | 0.9347 |
| R-2026 | 0.8865 | 0.7904 | 0.9645 | 0.8958 |
| R-2027 | 0.8254 | 0.7003 | 0.9477 | 0.9278 |
| R-2028 | 0.7475 | -0.3382 | 0.8812 | 0.8078 |
| R-2030 | 0.6478 | 0.5780 | 0.8774 | 0.8730 |
| R-2032 | 0.9409 | 0.8354 | 0.9705 | 0.9220 |
| R-2036 | 0.6662 | 0.8070 | 0.8217 | 0.9177 |
| R-2037 | 0.8222 | 0.3201 | 0.9114 | 0.7912 |
| R-2038 | 0.9145 | 0.8301 | 0.9587 | 0.9188 |
| R-2039 | 0.9521 | 0.2494 | 0.9760 | 0.8033 |
| R-2042 | 0.8903 | 0.8309 | 0.9500 | 0.9559 |
| R-2043 | 0.8322 | 0.8960 | 0.9237 | 0.9476 |

| Code | $NSE_{cal}$ | $NSE_{val}$ | $r^2_{cal}$ | $r^2_{val}$ |
|------|------|------|------|------|
| GS-3005 | 0.8307 | 0.5003 | 0.9156 | 0.8161 |
| GS-3016 | 0.8350 | 0.7798 | 0.9173 | 0.9126 |
| GS-3028 | 0.6290 | 0.6980 | 0.8320 | 0.8465 |
| GS-3035 | 0.5047 | -2.1931 | 0.8850 | 0.7974 |
| GS-3041 | 0.7106 | 0.1784 | 0.8497 | 0.8344 |
| GS-3047 | 0.6277 | 0.4813 | 0.8148 | 0.7479 |
| GS-3049 | 0.7333 | -0.0513 | 0.8738 | 0.8926 |
| GS-3051 | 0.7743 | 0.7591 | 0.8909 | 0.9226 |
| GS-3057 | 0.7230 | 0.6515 | 0.8864 | 0.9176 |
| GS-3089 | 0.9116 | 0.9026 | 0.9564 | 0.9561 |
| GS-3104 | 0.8137 | 0.6694 | 0.9208 | 0.8985 |
| GS-3105 | 0.6204 | 0.3723 | 0.8032 | 0.8010 |
| GS-3124 | 0.7397 | 0.3129 | 0.8604 | 0.8246 |
| GS-3150 | 0.7909 | 0.8507 | 0.9118 | 0.9436 |
| GS-3818 | 0.8724 | 0.8975 | 0.9403 | 0.9720 |

**Table 4.** Behavior of *NSE* in the Monte Carlo simulations for assessing equifinality and the efficiency of the calibration algorithm.

| Code | $NSE_{cal}$ | Number of simulations with $NSE > NSE_{cal} - 0.05$ | Number of simulations with $NSE > NSE_{cal}$ |
|------|-------------|----------------------------------------------------|---------------------------------------------|
| R-2001 | 0.8520 | 96 | 0 |
| R-2011 | 0.9237 | 1970 | 5 |
| R-2012 | 0.2723 | 426 | 0 |
| R-2013 | 0.8263 | 1259 | 0 |
| R-2014 | 0.8777 | 3463 | 0 |
| R-2026 | 0.8865 | 2804 | 8 |
| R-2027 | 0.8254 | 1863 | 56 |
| R-2028 | 0.7475 | 1364 | 2 |
| R-2030 | 0.6478 | 1132 | 0 |
| R-2032 | 0.9409 | 3568 | 0 |
| R-2036 | 0.6662 | 10 | 3 |
| R-2037 | 0.8222 | 0 | 0 |
| R-2038 | 0.9145 | 2808 | 0 |
| R-2039 | 0.9521 | 227 | 0 |
| R-2042 | 0.8903 | 368 | 0 |
| R-2043 | 0.8322 | 1660 | 5 |
| GS-3005 | 0.8307 | 32 | 5 |
| GS-3016 | 0.8350 | 1758 | 0 |
| GS-3028 | 0.6290 | 103 | 0 |
| GS-3035 | 0.5047 | 373 | 0 |
| GS-3041 | 0.7106 | 0 | 0 |
| GS-3047 | 0.6277 | 267 | 0 |
| GS-3049 | 0.7333 | 0 | 0 |
| GS-3051 | 0.7743 | 2446 | 0 |
| GS-3057 | 0.7230 | 518 | 3 |
| GS-3089 | 0.9116 | 3578 | 3 |
| GS-3104 | 0.8137 | 93 | 0 |
| GS-3105 | 0.6204 | 8 | 0 |
| GS-3124 | 0.7397 | 1 | 0 |
| GS-3150 | 0.7909 | 3484 | 0 |
| GS-3818 | 0.8724 | 26 | 0 |